\def\papertitle{Solid State Bus-Comp: A Large-Scale and Diverse Dataset for \\ Dynamic Range Compressor Virtual Analog Modeling}
\def\paperauthorA{Yicheng Gu}
\def\paperauthorB{Runsong Zhang$^*$}
\def\paperauthorC{Lauri Juvela}
\def\paperauthorD{Zhizheng Wu}

\documentclass[twoside,a4paper]{article}
\usepackage{etoolbox}
\usepackage{dafx_25}
\usepackage{amsmath,amssymb,amsfonts,amsthm}
\usepackage{euscript}
\usepackage[T1]{fontenc}
\usepackage[utf8]{inputenc}
\usepackage{ifpdf}
\usepackage[english]{babel}
\usepackage{caption}
\usepackage{color}
\usepackage{makecell}
\usepackage{url}
\usepackage{booktabs} 
\usepackage{nicefrac}
\usepackage{microtype}
\usepackage{xcolor}
\usepackage{multirow}
\usepackage{algorithmic}
\usepackage{array}
\usepackage{textcomp}
\usepackage{stfloats}
\usepackage{verbatim}
\usepackage{subcaption}
\usepackage{mathptmx}
\usepackage{tikz}
\usepackage{pgf-pie}
\usepackage{lipsum}

\input glyphtounicode
\makeatletter
\newcommand\footnoteref[1]{\protected@xdef\@thefnmark{\ref{#1}}\@footnotemark}
\makeatother

\ninept

\newcounter{numauth}
\newcounter{listcnt}
\newcommand\authcnt[1]{\ifdefined#1 \stepcounter{numauth} \fi}

\newcommand\addauth[1]{
\ifdefined#1 
\stepcounter{listcnt}
\ifnum \value{listcnt}<\value{numauth}
\appto\authorslist{, #1}
\else
\appto\authorslist{~and~#1}
\fi
\fi}
\authcnt{\paperauthorB}
\authcnt{\paperauthorC}
\authcnt{\paperauthorD}
\authcnt{\paperauthorE}
\authcnt{\paperauthorF}
\authcnt{\paperauthorG}
\authcnt{\paperauthorH}
\authcnt{\paperauthorI}
\authcnt{\paperauthorJ}
\def\authorslist{\paperauthorA}
\addauth{\paperauthorB}
\addauth{\paperauthorC}
\addauth{\paperauthorD}
\addauth{\paperauthorE}
\addauth{\paperauthorF}
\addauth{\paperauthorG}
\addauth{\paperauthorH}
\addauth{\paperauthorI}
\addauth{\paperauthorJ}

\usepackage{times}

\newif\ifpdf
\ifx\pdfoutput\relax
\else
   \ifcase\pdfoutput
      \pdffalse
   \else
      \pdftrue
   \fi
\fi

\ifpdf 
  \usepackage[pdftex,
    pdftitle={\papertitle},
    pdfauthor={\authorslist},
    pdfsubject={Proceedings of the 28th International Conference on Digital Audio Effects (DAFx25)},
    colorlinks=false, 
    bookmarksnumbered, 
    pdfstartview=XYZ 
  ]{hyperref}
\else 
  \usepackage[dvips,
    pdftitle={\papertitle},
    pdfauthor={\authorslist},
    pdfsubject={Proceedings of the 28th International Conference on Digital Audio Effects (DAFx25)},
    colorlinks=false, 
    bookmarksnumbered, 
    pdfstartview=XYZ 
  ]{hyperref}
\fi
\usepackage[hypcap=true]{caption}
\title{\papertitle}

\affiliation{
\paperauthorA \sthanks{Equal Contribution}$^{1}$$^{2}$,
\paperauthorB $^{2}$,
\paperauthorC $^{1}$ and 
\paperauthorD $^{2}$
}
{
$^1$\href{https://www.aalto.fi/en/aalto-acoustics-lab}{Acoustics Lab}, Aalto University, Espoo, Finland \\
$^2${\href{https://sds.cuhk.edu.cn/en}{School of Data Science}, The Chinese University of Hong Kong, Shenzhen, China  \\
}
{\tt \href{mailto:yicheng.gu@aalto.fi}{yicheng.gu@aalto.fi}}
}

\begin{document}
\ifpdf 
  \DeclareGraphicsExtensions{.png,.jpg,.pdf}
\else  
  \DeclareGraphicsExtensions{.eps}
\fi


\maketitle

\begin{abstract}

Virtual Analog (VA) modeling aims to simulate the behavior of hardware circuits via algorithms to replicate their tone digitally. Dynamic Range Compressor (DRC) is an audio processing module that controls the dynamics of a track by reducing and amplifying the volumes of loud and quiet sounds, which is essential in music production. In recent years, neural-network-based VA modeling has shown great potential in producing high-fidelity models. However, due to the lack of data quantity and diversity, their generalization ability in different parameter settings and input sounds is still limited. To tackle this problem, we present Solid State Bus-Comp, the first large-scale and diverse dataset for modeling the classical VCA compressor --- SSL 500 G-Bus. Specifically, we manually collected 175 unmastered songs from the Cambridge Multitrack Library. We recorded the compressed audio in 220 parameter combinations, resulting in an extensive 2528-hour dataset with diverse genres, instruments, tempos, and keys. Moreover, to facilitate the use of our proposed dataset, we conducted benchmark experiments in various open-sourced black-box and grey-box models, as well as white-box plugins. We also conducted ablation studies in different data subsets to illustrate the effectiveness of the improved data diversity and quantity. The dataset and demos are on our project page: \url{https://www.yichenggu.com/SolidStateBusComp/}.

\end{abstract}

\section{Introduction}
\label{sec:intro}

Virtual Analog (VA) modeling aims to simulate analog audio devices digitally. Dynamic Range Compressor (DRC) is an audio processing module that compresses the dynamics of a track by reducing and amplifying the volumes of loud and quiet sounds, which is essential in music production~\cite{drc-review-2012}. VA modeling on DRC is important, but is always considered to be challenging due to its characteristics: non-linear and long temporal dependency.

To model an analog compressor, early DSP-based methods utilized white-box models. Such a model generally comprises a gain computer and a level detector with different algorithm designs~\cite{review-1984,drc-review-2012}, which have been well-studied over the years. Apart from this, recent works have also been proposed to explore other potential improvements like increasing computational efficiency~\cite{block-drc, FFT-drc} and integrating machine-learning techniques for automatic mixing~\cite{automatic-2013,multitrack-2013}. These developments have led to various achievements in modeling both the entire device~\cite{device-2011} and specific components~\cite{optocoupler-2016,data-2023}.

Although these white-box techniques can deliver high-quality modeling over different devices, the involvement of human experts is often needed, making it hard to automate the modeling process. In recent years, neural-network-based black-box models have developed a lot due to their superior ability to model analog devices in a data-driven way. To be specific, ~\cite{la2a} first proposed an autoencoder model to model various audio effects. ~\cite{LSTM-2019} utilized the long short-term memory (LSTM) model for optimizing the long-term dependencies, followed by~\cite{hrnn} to further expand into the hyper recurrent neural network (RNN) model with an in-depth comparison between RNN and LSTM models. To utilize the advantages of convolutional neural networks (CNNs), ~\cite{wavenet-audio} first employed the WaveNet~\cite{wavenet} structure on digital audio effects. Based on this work, ~\cite{TCN-2021} proposed a temporal convolutional network (TCN) with larger receptive fields and huge dilation factors, while ~\cite{GCN-2023} further improved this architecture by integrating the feature-wise linear modulation (FiLM)~\cite{film} layers in modeling the parameter conditions. State Space Model (SSM)~\cite{ssm-2022} is another technique to model long-term dependent time series via decomposing a dynamic system into structured state variables. ~\cite{S4-2024} first employed the S4 blocks in VA modeling, obtaining outstanding performances, followed by ~\cite{S6-2024} further adopting the latest S6 model~\cite{mamba}. With the development of differentiable digital signal processing (DDSP)~\cite{ddsp}, works are also proposed to integrate the DSP models' explainability and efficiency with neural networks. For instance, ~\cite{biquad-2020} proposed differentiable biquad filters for deep learning applications, followed by~\cite{klann-2024} integrating them with Koopman Networks~\cite{koopman-network} to operate in a higher-dimensional state space. These advances have also made the neural grey-box models viable. In particular, ~\cite{DRC-2022} utilized the classic white-box DRC~\cite{drc-review-2012} design with multilayer perceptrons (MLPs) predicting the parameters in each time frame, followed by ~\cite{nablafx} to further simplify the model into parametric Gains for compression and supplementary EQs for non-linear distortion.

Despite the rapid development of VA models, the publicly available datasets are still scarce, with limited data quantity and diversity. Table~\ref{tab:existing_datasets} illustrates the details of the existing datasets regarding DRC. Specifically, early attempts~\cite{UA6176} primarily consist of processed short instrument and test signal recordings in a specific parameter setting, tailored for trivial non-parameteric models. SignalTrain~\cite{la2a} first proposed a parametric dataset in modeling the optical compressor LA-2A. It used various randomly generated test signals and a few instrument recordings as the input signals and recorded 20 equally sampled parameter combinations. After that, ~\cite{cl1b} proposed the CL-1B dataset with real-world recordings as inputs with more parameter combinations. Recent works like~\cite{S6-2024} also presented datasets with more diverse devices but often with limited data scale and parameter combinations. Such limitations will significantly constrain the model's performance, especially when encountering real-world recordings and unseen parameters.

Data scaling has been shown to be effective in many audio-related areas~\cite{mert, stableaudio, maskgct, singnet, neurodyne}. For instance, Mert~\cite{mert} utilized a music mixture of 160K hours to scale up a self-supervised representation learning model with 330M parameters, obtaining outstanding performance in music information retrieval; Yue~\cite{yue} constructed a 650K hours music mixture to train a 7B parameter model for music generation, obtaining state-of-the-art (SOTA) performance; Stable Audio~\cite{stableaudio} collected 73k hours of audio recordings, leading to SOTA audio generation model with 1B parameters; Emilia~\cite{emilia, emilia-journal} presented a 101K hours open-sourced speech dataset, facilitating SOTA speech generations models~\cite{maskgct, vevo}.

Following these previous works, this work presents Solid State Bus-Comp, the first large-scale and diverse dataset for modeling the SSL 500 G-Bus Compressor~\footnote{\label{sslgbus}\url{https://solidstatelogic.com/products/stereo-bus-compressor-module}}. Specifically, we manually selected 175 unmastered real-world songs from the Cambridge Multitrack Library~\footnote{\label{cambridge}\url{https://www.cambridge-mt.com/ms/mtk/}} and recorded the compressed signals in 220 parameter combinations, which results in an extensive 2528-hour dataset with diverse genres, instruments, tempos, and keys. To facilitate the use of our dataset, we conducted benchmarking experiments on various open-sourced black-box and grey-box models, as well as available white-box plugins. We also conducted ablation studies on data subsets with different amounts of songs and data scales to illustrate the effectiveness of the improved data diversity and quantity.

\begin{table*}[t]
    \centering
    \caption{A comparison of Solid State Bus-Comp with existing VA modeling datasets regarding DRC.}
    \label{tab:existing_datasets}
        \begin{tabular}{cccccc}
            \toprule
            \textbf{Device} & \textbf{Duration (hour)} & \textbf{Type} & \textbf{Parameters} & \textbf{Range} & \textbf{Combinations} \\
            \midrule
            \makecell{UA 6176 \\ Limiter \\ \cite{UA6176}} & 0.66 & \makecell{Transistor-Based \\ Limiter} & \makecell{Attack \\ Release \\ Input Level \\ Output Level \\ Ratio} & \makecell{800\,$\mu\text{s}$ \\ 1100\,ms \\ 4 \\ 7 \\ All} & 1 \\
            \midrule
            \makecell{Ampeg \\ Opto Comp \\ \cite{tonetwist}}  & 3.61 & \makecell{Optical \\ Compressor} & \makecell{Compression \\ Release \\ Level} & \makecell{\textnormal{[}3, 10\textnormal{]} \\ \textnormal{[}1, 10\textnormal{]}\,s \\ 6} & 5 \\
            \midrule
            \makecell{Flamma \\ FC21 \\ \cite{tonetwist}}  & 3.61 & \makecell{Optical \\ Compressor} & \makecell{Comp \\ EQ \\ Volumn} & \makecell{\textnormal{[}1, 10\textnormal{]} \\ \textnormal{[}1, 10\textnormal{]} \\ 10} & 5 \\
            \midrule
            \makecell{Yuer \\ RF-10 \\ \cite{tonetwist}}  & 3.61 & \makecell{OTA \\ Compressor} & \makecell{ Attack \\ Sustain \\ level} & \makecell{\textnormal{[}1, 10\textnormal{]}\,ms \\ \textnormal{[}1, 10\textnormal{]}\,ms \\ 10} & 6 \\
            \midrule
            \makecell{Teletronix \\ LA-2A \\ \cite{la2a}}  & 48.63 & \makecell{Optical \\ Compressor} & \makecell{ Peak Reduction \\ Switch Mode} & \makecell{\textnormal{[}0, 100\textnormal{]} \\ \textnormal{[}Compressor, Limiter\textnormal{]}} & 20 \\
            \midrule
            \makecell{TubeTech \\ CL-1B \\ \cite{cl1b}}  & 37.54 & \makecell{Optical \\ Compressor} & \makecell{Threshold \\ Attack \\ Release \\ Ratio} & \makecell{\textnormal{[}-40, 0\textnormal{]}\,dB \\ \textnormal{[}5, 300\textnormal{]}\,ms \\ \textnormal{[}0.005, 10\textnormal{]}\,s \\ 1:\textnormal{[}1, 10\textnormal{]}} & 108 \\
            \midrule
            \makecell{SSL 500 \\ G-Bus-Comp \\ (ours)} & 2528.53 & \makecell{VCA \\ Compressor} & \makecell{Threshold \\ Attack \\ Release \\ Ratio} & \makecell{\textnormal{[}-40, 0\textnormal{]}\,dB \\ \textnormal{[}0.1, 30\textnormal{]}\,ms \\ \textnormal{[}0.1, 1.6\textnormal{]}\,s \\ 1:\textnormal{[}1.5, 10\textnormal{]}} & 220 \\
            \bottomrule
        \end{tabular}
\end{table*}

\section{Solid State Bus-Comp}
\label{sec:ssl-bus}

\begin{figure*}[t]
    \centering
    \begin{subfigure}[b]{0.45\textwidth}
    \centering
        \begin{tikzpicture}
            \definecolor{color1}{HTML}{447cac}
            \definecolor{color2}{HTML}{88ce9b}
            \definecolor{color3}{HTML}{e3f79b}
            \definecolor{color4}{HTML}{fae28c}
            \definecolor{color5}{HTML}{f1874b}
            \definecolor{color6}{HTML}{c42d40}
            \definecolor{color7}{HTML}{7695FF}
            \definecolor{color8}{HTML}{9DBDFF}
            \pie[
                text=legend,
                radius=2,
                color={color1, color2, color3, color4, color5, color6, color7, color8},
                explode=0.1, 
                sum=auto, 
                before number=\phantom,
                after number=
            ]{
                5.33/Rock: 5.33,
                4.45/Pop: 4.45, 
                2.79/Electronic: 2.79, 
                1.87/Folk: 1.87,
                1.17/Ambient: 1.17,
                1.03/Metal: 1.03,
                0.56/Jazz: 0.56, 
                0.20/Hip Hop: 0.20
            }
        \end{tikzpicture}
    \caption{Genre}
    \label{fig:genre}
    \end{subfigure} 
    \hfill
    \begin{subfigure}[b]{0.45\textwidth}
    \centering
        \begin{tikzpicture}
            \definecolor{color1}{HTML}{447cac}
            \definecolor{color2}{HTML}{88ce9b}
            \definecolor{color3}{HTML}{e3f79b}
            \definecolor{color4}{HTML}{fae28c}
            \definecolor{color5}{HTML}{f1874b}
            \definecolor{color6}{HTML}{c42d40}
            \definecolor{color7}{HTML}{7695FF}
            \definecolor{color8}{HTML}{9DBDFF}
            \definecolor{color9}{HTML}{FFD7C4}
            \definecolor{color10}{HTML}{884EA0}
            \pie[
                text=legend,
                radius=2,
                color={color1, color2, color3, color4, color5, color6, color7, color8, color9, color10},
                explode=0.1, 
                sum=auto, 
                before number=\phantom,
                after number=
            ]{
                11.43/Bass: 11.43,
                11.43/Drum: 11.43,
                11.02/Guitar: 11.02, 
                9.73/Vocal: 9.73,
                9.24/Synth: 9.24,
                6.59/Piano: 6.59, 
                5.06/String: 5.06, 
                3.40/Keyboard: 3.40,
                2.48/Brass: 2.48,
                0.50/Woodwind: 0.50
            }
        \end{tikzpicture}
    \caption{Instrument}
    \label{fig:instruments}
    \end{subfigure} 
    \caption{Duration statistics (hours) of the unmastered songs used as input signals in Solid State Bus-Comp by genres and instruments. }
    \label{fig:statistics}
    \vspace{-5pt}
\end{figure*}
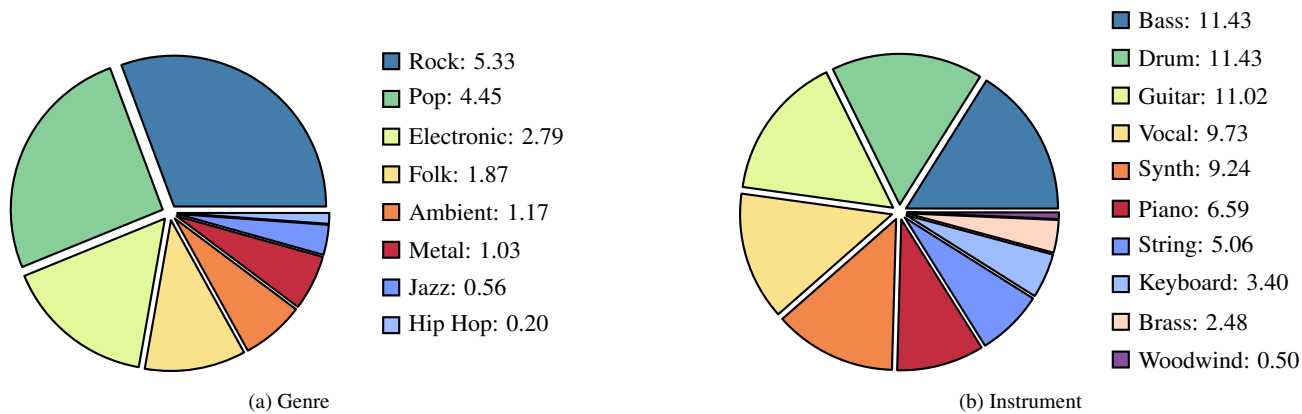

This section provides the construction details, statistics, and analysis of our proposed Solid State Bus-Comp dataset.

\subsection{Dataset Construction}

Solid State Bus-Comp comprises unmastered songs with different genres, instruments, tempos, and keys processed with varying compression parameters. In particular, we manually selected 175 unmastered songs from the Cambridge Multitrack Library~\footnoteref{cambridge}. We used Reaper~\footnote{\url{https://www.reaper.fm/}} as the Digital Audio Workstation (DAW) to process the data automatically. Specifically, we used the RME Fireface UFX+~\footnote{\url{https://rme-audio.de/fireface-ufx.html}} as the external audio interface and connected it to the ReaInsert. Then, we wrote a ReaScript to automatically send and receive signals from the hardware compressor via the audio interface. To match the level between the DAW and hardware compressor, we normalized all songs to -12\,dB and applied a 5\,dB input boost and a 5\,dB output attenuation. We manually selected 144 widely used parameter combinations for processing after consulting six professional mastering engineers, which are: threshold [-28, -24, -20, -16], attack [0.1, 0.3, 1, 3], release [0.1, 0.4, 0.8, auto], ratio [2, 4, 10]. We additionally recorded 76 other randomly selected combinations as supplementary edge cases. All the audio data was recorded at a sampling rate of 44.1\,kHz.

\begin{figure}[ht]
    \centering
    \resizebox{\linewidth}{!}{
        \includegraphics[]{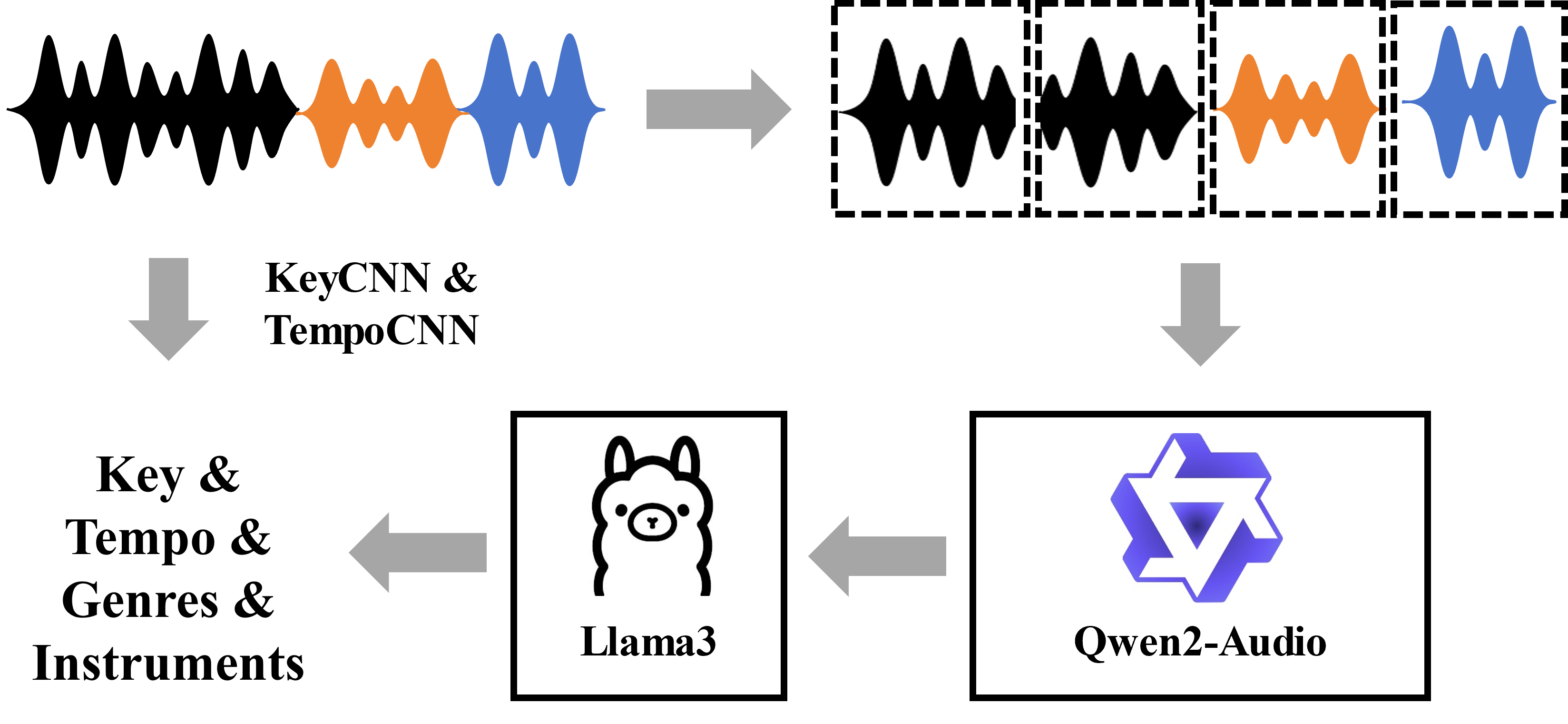}
    }
    \caption{The annotation pipeline of Solid State Bus-Comp. We utilized various pre-trained models to obtain information on each song's key, tempo, genre, and instrument.}
    \label{fig:annotation}
\vspace{-5pt}
\end{figure}

\subsection{Dataset Statistics}

We utilized various pre-trained models to annotate our data, as illustrated in Fig.~\ref{fig:annotation}. Specifically, we used the KeyCNN~\footnote{\url{https://github.com/hendriks73/key-cnn}} and TempoCNN model~\footnote{\url{https://github.com/hendriks73/tempo-cnn}} proposed in~\cite{tempocnn}~\footnote{\url{https://github.com/hendriks73/directional_cnns}} to obtain the global music tempo and key information. We split each song into a series of 10s segments and used the Qwen2-Audio~\cite{qwen2audio}~\footnote{\url{https://huggingface.co/Qwen/Qwen2-Audio-7B}} to annotate each segment's content, which will then be fed to a Llama3~\cite{llama3}~\footnote{\url{https://huggingface.co/meta-llama}} model to organize the genres and instruments of the whole song. 

\begin{figure*}[t]
    \centering
    \begin{subfigure}[b]{0.48\textwidth}
    \centering
    \includegraphics[width=\textwidth]{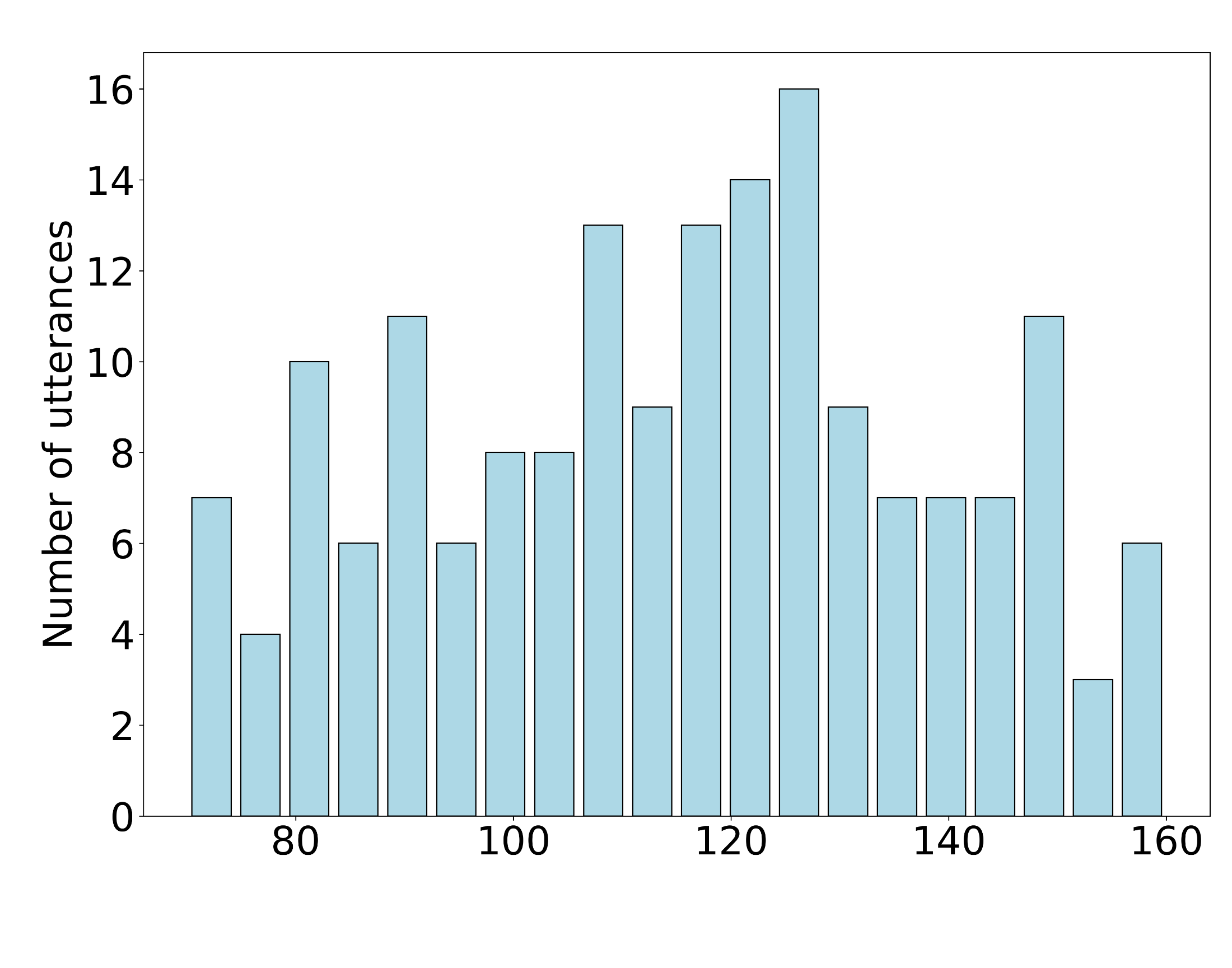}
     \caption{Tempo}
     \label{fig:tempo}
    \end{subfigure} 
    \hfill
    \begin{subfigure}[b]{0.48\textwidth}
    \centering
    \includegraphics[width=\textwidth]{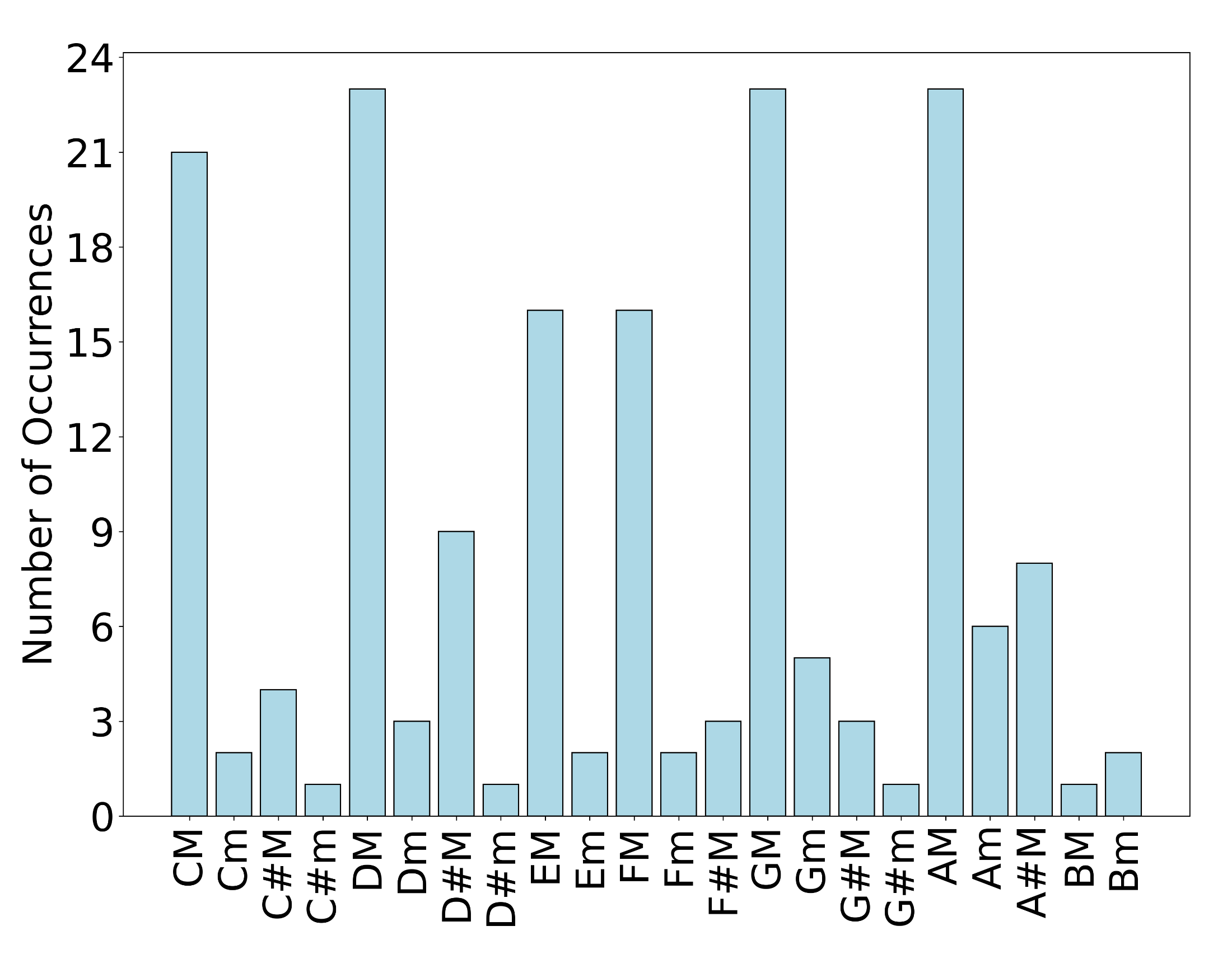}
     \caption{Key}
     \label{fig:mode}
    \end{subfigure} 
    \caption{Tempo and Key statistics (occurrences) of the unmastered songs used as input signals in our proposed Solid State Bus-Comp. Tempo is in beats per minute (BPM). ``M'' denotes for ``Major'' and ``m'' denotes for ``Minor''.}
    \label{fig:statistics-second}
\end{figure*}

The statistical results of Solid State Bus-Comp on genres, instruments, tempos, and keys are illustrated in Fig.~\ref{fig:statistics} and Fig.~\ref{fig:statistics-second}. From these results, we can conclude that 1) The majority of genres in our dataset are Rock, Pop, Electronic, and Folk, with a small amount of other uncommon ones like Ambient, Metal, Jazz and Hip Hop; 2) Most used instruments in our dataset are Bass, Drum, Guitar, Vocal, and Synth, with a considerable amount of Piano, String, Keyboard, and Brass. Niche instruments, like Woodwind, are also presented in the dataset; 3) Songs in our dataset are within the range of 70-160 beats per minute (BPM), and the majority of songs are distributed around 110-130 BPM; 4) Most songs in our dataset are in C, D, E, F, G, and A Majors, with a small number of remaining songs evenly distributed across other keys.

\begin{figure*}[t]
    \centering
    \begin{subfigure}[b]{0.48\textwidth}
    \centering
    \includegraphics[width=\textwidth,height=\textwidth]{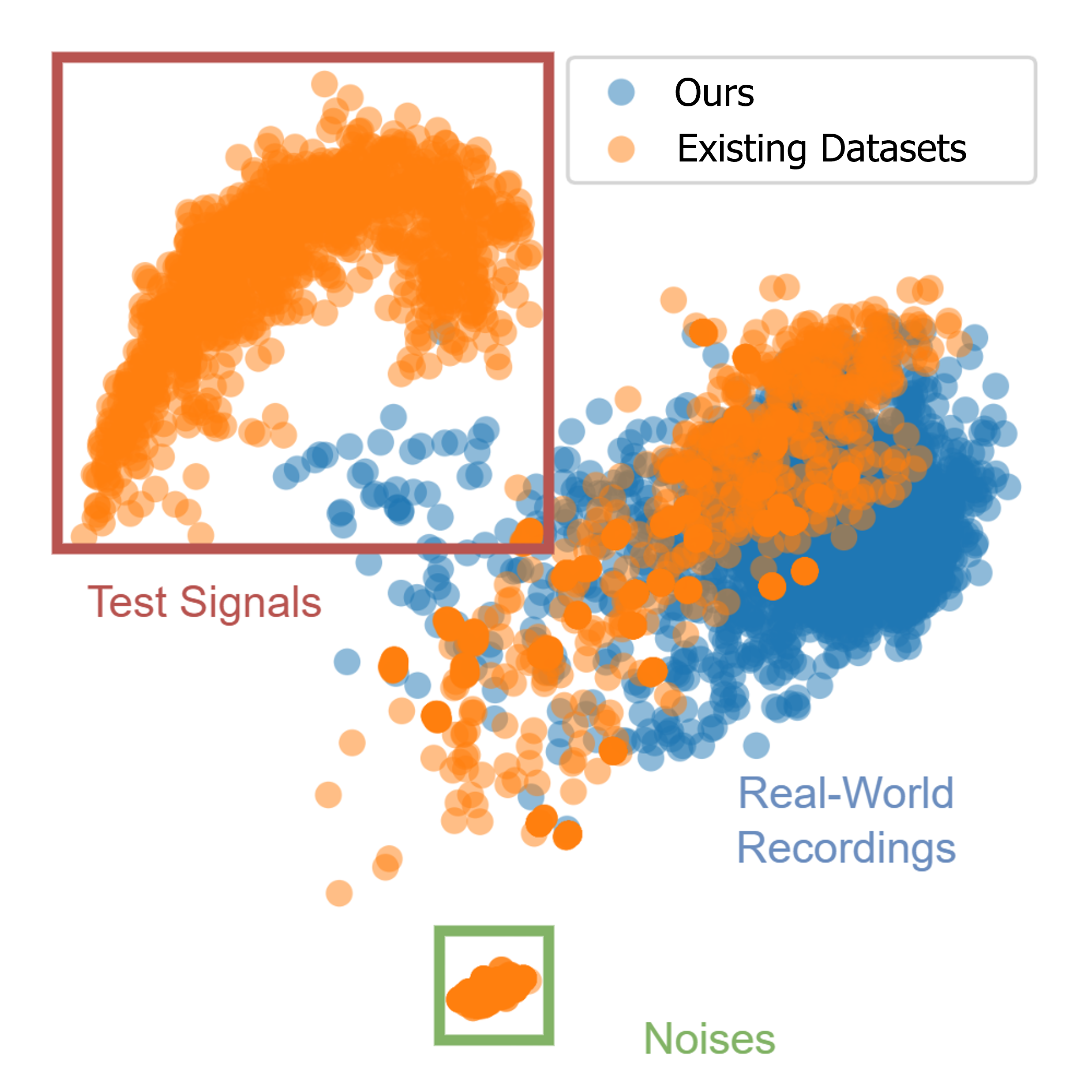}
     \caption{Acoustic Diversity}
     \label{fig:pca-acoustic}
    \end{subfigure} 
    \hfill
    \begin{subfigure}[b]{0.48\textwidth}
    \centering
    \includegraphics[width=\textwidth,height=\textwidth]{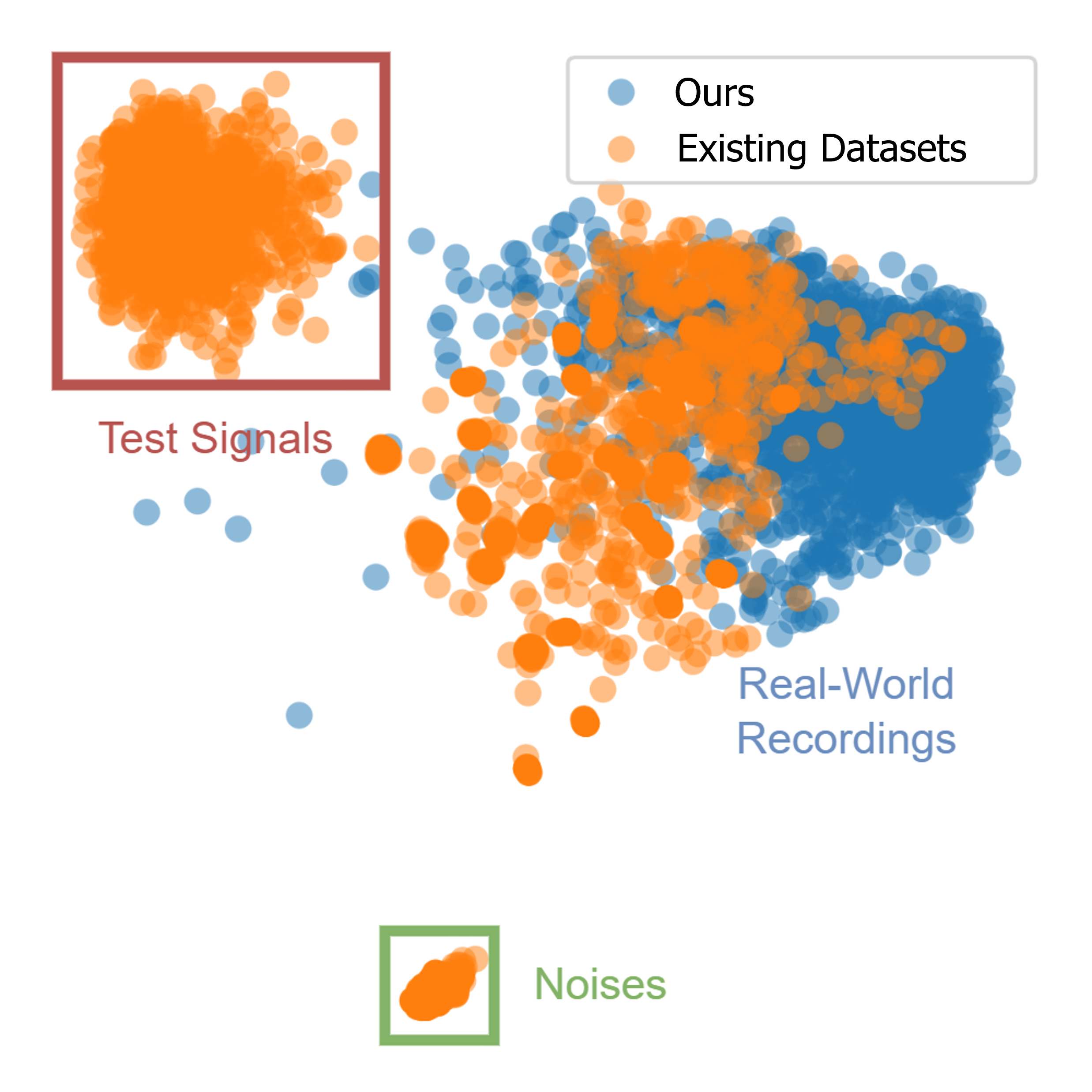}
     \caption{Semantic Diversity}
     \label{fig:pca-semantic}
    \end{subfigure} 
    \caption{Comparison of acoustic and semantic diversities in input signals between Solid State Bus-Comp and the existing datasets. The plottings are obtained by applying the PCA algorithm to the SSL representations. We used MERT to extract acoustic embeddings and w2v-BERT 2.0 to extract semantic embeddings. For existing datasets, the compact cluster represents random noises, the diffused cluster represents test signals (sine, square, triangle waves, etc), and the remaining scattered points represent real-world recordings.}
    \label{fig:pca}
\end{figure*}

\subsection{Dataset Analysis}

Unlike existing datasets, which primarily utilize noises and analysis signals, Solid State Bus-Comp comprises a collection of diversified real-world unmastered songs as the input signals. To quantify this diversity, we use self-supervised learning (SSL) models to investigate and compare their differences in acoustic and semantic feature spaces, following~\cite{emilia},~\cite{emilia-journal}, and~\cite{singnet}. 

Specifically, to analyze the diversity of acoustic features, we leveraged a pre-trained MERT~\cite{mert}~\footnote{\url{https://huggingface.co/m-a-p/MERT-v1-330M}} model to extract the acoustic representation (the 12th layer is used), which captures various acoustic characteristics such as timbre, style, key, etc. For the semantic diversity analysis, we employed a pre-trained w2v-BERT model~\cite{w2vbert}~\footnote{\url{https://huggingface.co/facebook/w2v-bert-2.0}} to generate semantic representations (the last layer is used), capturing melody, lyrics, rhythm, etc. We then applied the Principal Component Analysis (PCA) algorithm to reduce the dimensionality of these representations to two. As illustrated in Fig.~\ref{fig:pca}, most sample points in existing datasets are centered in two distant clusters, where the compact one represents the noise signals, and the diffused one represents the test signals (sine, square, triangle waves, and their combinations), and only a few points scattered aside, representing the real-world instrument recordings. Compared with the existing datasets, Solid State Bus-Comp exhibits a broader dispersion in the cluster representing real-world recordings, indicating richer acoustic and semantic characteristic coverage. 

\section{Experiments}

\begin{table*}[t]
\begin{center}
\caption{Benchmarking results of existing parametric black-box methods. The best and second best results are \textbf{bold} and \underline{underlined}.}
\label{tab:results-black-box}
\begin{tabular}{lccccccc}

\toprule
\multirow{2}{*}{\textbf{System}} & \multirow{2}{*}{\textbf{Configuration}} & \multirow{2}{*}{\textbf{Condition}} & \multirow{2}{*}{\textbf{\#Params}} & \multicolumn{2}{c}{\textbf{L1 ($\downarrow$)}} & \multicolumn{2}{c}{\textbf{M-STFT ($\downarrow$)}} \\ 
\cmidrule(lr){5-6} \cmidrule(lr){7-8}
& & & &\textbf{Seen} & \textbf{Unseen} & \textbf{Seen} & \textbf{Unseen} \\
\midrule
\multirow{4}{*}{LSTM~\cite{LSTM-2019}} & \multirow{2}{*}{32 Channels} & Concat & 5.0K & 0.0290 & 0.0239 & 0.3954 & 0.4644 \\
 & & TVConcat & 8.0K & 
 \underline{0.0030} & \underline{0.0028} & 0.3631 & 0.4523 \\
\cmidrule(lr){2-8}
 & \multirow{2}{*}{96 Channels} & Concat & 39.7K & 0.0274 & 0.0237 & 0.4732 & 0.8123 \\
 & & TVConcat & 45.7K & \textbf{0.0028} & 0.0029 & 0.4256 & 0.5483 \\
\midrule
\multirow{8}{*}{TCN~\cite{TCN-2021}} & \multirow{4}{*}{\makecell{5 Blocks \\ 7 Kernel \\ 4 Dilation}} & FiLM & 15.0K & 0.0296 & 0.0251 & 0.5432 & 0.8647 \\
 & & TFiLM & 42.0K & 0.0066 & 0.0056 & 0.3755 & 0.4492 \\
 & & TTFiLM & 17.3K & 0.0271 & 0.0224 & 0.3903 & 0.4953 \\
 & & TVFiLM & 17.7K & 0.0252 & 0.0224 & 0.5957 & 0.9704 \\
 \cmidrule(lr){2-8}
 & \multirow{4}{*}{\makecell{10 Blocks \\ 3 Kernel \\ 2 Dilation}} & FiLM & 20.1K & 0.0088 & 0.0079 & 0.5158 & 0.6959 \\
 & & TFiLM & 76.4K & 0.0080 & 0.0067 & 0.3731 & 0.4427 \\
 & & TTFiLM & 27.0K & 0.0260 & 0.0215 & 0.3804 & 0.5057 \\
 & & TVFiLM & 22.8K & 0.0083 & 0.0069 & 0.3819 & \textbf{0.3983} \\
\midrule
\multirow{8}{*}{GCN~\cite{GCN-2023}} & \multirow{4}{*}{\makecell{5 Blocks \\ 7 Kernel \\ 4 Dilation}} & FiLM & 29.0K & 0.0271 & 0.0223 & 0.4760 & 0.5527 \\
 & & TFiLM & 146.0K & 0.0041 & 0.0034 & 0.3713 & \underline{0.4045} \\
 & & TTFiLM & 31.6K & 0.0066 & \textbf{0.0024} & 0.3817 & 0.5766 \\
 & & TVFiLM & 31.7K & 0.0270 & 0.0226 & 0.3406 & 0.4147 \\
 \cmidrule(lr){2-8}
 & \multirow{4}{*}{\makecell{10 Blocks \\ 3 Kernel \\ 2 Dilation}} & FiLM & 40.5K & 0.0241 & 0.0200 & 0.6757 & 0.6346 \\
 & & TFiLM & 278.0K & 0.0267 & 0.0220 & 0.3497 & 0.4438 \\
 & & TTFiLM & 48.0K & 0.0063 & \textbf{0.0024} & 0.3549 & 0.5766 \\
 & & TVFiLM & 43.2K & 0.0272 & 0.0226 & \textbf{0.3238} & 0.4456 \\
\midrule
\multirow{8}{*}{S4~\cite{S4-2024}} & \multirow{4}{*}{\makecell{4 Blocks \\ 4 State Dimension}} & FiLM & 8.9K & 0.0287 & 0.0246 & 0.8044 & 1.0532 \\
 & & TFiLM & 30.0K & 0.0277 & 0.0230 & 0.3576 & 0.4973 \\
 & & TTFiLM & 10.2K & \underline{0.0030} & 0.0030 & 0.3884 & 0.4689 \\
 & & TVFiLM & 11.6K & 0.0283 & 0.0237 & 0.3898 & 0.5842 \\
 \cmidrule(lr){2-8}
 & \multirow{4}{*}{\makecell{8 Blocks \\ 32 State Dimension}} & FiLM & 29.7K & 0.0103 & 0.0102 & 1.0552 & 1.2474 \\
 & & TFiLM & 74.3K & 0.0046 & 0.0043 & 0.4961 & 0.6098 \\
 & & TTFiLM & 34.8K & 0.0265 & 0.0225 & 0.4665 & 0.5898 \\
 & & TVFiLM & 32.4K & \underline{0.0030} & 0.0031 & \underline{0.3480} & 0.4930 \\
\bottomrule

\end{tabular}
\end{center}
\vspace{-15pt}
\end{table*}

\begin{table*}[t]
\begin{center}
\caption{Benchmarking results of existing grey-box methods. The best and second best results are \textbf{bold} and \underline{underlined}.}
\label{tab:results-grey-box}
\begin{tabular}{lcccccccc}

\toprule
\multirow{2}{*}{\textbf{System}} & \multicolumn{3}{c}{\textbf{Signal Chain}} & \multirow{2}{*}{\textbf{\#Params}} & \multicolumn{2}{c}{\textbf{L1 ($\downarrow$)}} & \multicolumn{2}{c}{\textbf{M-STFT ($\downarrow$)}} \\ 
\cmidrule(lr){2-4} \cmidrule(lr){6-7} \cmidrule(lr){8-9}
& \textbf{Static Gain} & \textbf{Make-Up Gain} & \textbf{Level Detector} & &\textbf{Seen} & \textbf{Unseen} & \textbf{Seen} & \textbf{Unseen} \\
\midrule
\multirow{6}{*}{GreyBoxDRC~\cite{DRC-2022}} & \multirow{3}{*}{Soft Knee} & \multirow{3}{*}{Static Gain} & One-Pole & 0.6K & 0.0076 & 0.0066 & 1.0046 & 1.1312 \\
 & & &  Switching One-Pole & 0.6K & 0.0067 & 0.0072 & \underline{0.8108} & 1.2388 \\
 & & &  RNN Mod. One-Pole & 0.7K & 0.0076 & 0.0070 & 1.0066 & 1.2251 \\
 \cmidrule{2-9}
 & \multirow{3}{*}{Hard Knee} & \multirow{3}{*}{GRU} & One-Pole & 0.8K & 0.0062 & 0.0074 & 1.1072 & 1.5134 \\
 & & &  Switching One-Pole & 0.8K & \underline{0.0059} & \underline{0.0061} & 0.8758 & \underline{1.0888} \\
 & & &  RNN Mod. One-Pole & 0.9K & 0.0061 & 0.0070 & 1.1218 & 1.6492 \\
\midrule
\multirow{2}{*}{ToneTwist~\cite{tonetwist}} & \multicolumn{3}{c}{PEQ $\rightarrow$ Gain $\rightarrow$ PEQ $\rightarrow$ Gain} & 1.6K & \textbf{0.0034} & \textbf{0.0034} & \textbf{0.4098} & \textbf{0.6004} \\
 & \multicolumn{3}{c}{PEQ $\rightarrow$ Phase Inversion $\rightarrow$ Gain $\rightarrow$ PEQ $\rightarrow$ Gain} & 2.0K & 0.0200 & 0.0168 & 1.5964 & 1.4596 \\
\bottomrule

\end{tabular}
\end{center}
\vspace{-20pt}
\end{table*}

\begin{table}[t]
\begin{center}
\caption{Benchmarking results of existing commercial plugins. The best and second best results are \textbf{bold} and \underline{underlined}.}
\label{tab:results-plugins}
\begin{tabular}{lcccc}

\toprule
\multirow{2}{*}{\textbf{System}} & \multicolumn{2}{c}{\textbf{L1 ($\downarrow$)}} & \multicolumn{2}{c}{\textbf{M-STFT ($\downarrow$)}} \\ 
\cmidrule(lr){2-3} \cmidrule(lr){4-5}
& \textbf{Seen} & \textbf{Unseen} & \textbf{Seen} & \textbf{Unseen} \\
\midrule
Solid State Logic & \underline{0.0322} & \underline{0.0175} & \underline{0.4489} & \underline{0.2943} \\
Softube & 0.0448 & 0.0237 & 0.7069 & 0.4546 \\
Overloud & 0.0326 & 0.0176 & 0.4738 & 0.3253 \\ 
PSPaudioware & \textbf{0.0269} & \textbf{0.0145} & \textbf{0.3047} & \textbf{0.2184} \\
\bottomrule

\end{tabular}
\end{center}
\vspace{-10pt}
\end{table}

\begin{table}[t]
\begin{center}
\caption{Ablation results of the GCN model trained on different data subsets. The best and second best results of every column in each setting are \textbf{bold} and \underline{underlined}.}
\label{tab:results-ablation}
\begin{tabular}{cccccc}

\toprule
\multirow{2}{*}{\textbf{\#Songs}} & \multirow{2}{*}{\makecell{\textbf{Duration} \\ \textbf{(hour)}}} & \multicolumn{2}{c}{\textbf{L1 ($\downarrow$)}} & \multicolumn{2}{c}{\textbf{M-STFT ($\downarrow$)}} \\ 
\cmidrule(lr){3-4} \cmidrule(lr){5-6}
& & \textbf{Seen} & \textbf{Unseen} & \textbf{Seen} & \textbf{Unseen} \\
\midrule
\multirow{5}{*}{100} & 0.05 & 0.0279 & 0.0262 & 0.4718 & 0.5699 \\
 & 0.5 & 0.0278 & \underline{0.0226} & 0.3649 & 0.4838 \\
 & 5 & 0.0298 & \textbf{0.0222} & 0.3641 & 0.4539 \\
 & 50 & \underline{0.0273} & 0.0227 & \underline{0.3245} & \underline{0.4520} \\
 & 500 & \textbf{0.0272} & \underline{0.0226} & \textbf{0.3238} & \textbf{0.4456} \\
\midrule
5 & \multirow{5}{*}{50} & 0.0276 & 0.0231 & 0.4793 & 0.5876 \\
10 & & 0.0277 & 0.0230 & 0.4294 & 0.5159 \\
25 & & 0.0277 & \underline{0.0227} & 0.3333 & 0.5030 \\
50 &  & \underline{0.0274} & \textbf{0.0224} & \underline{0.3247} & \textbf{0.4502} \\ 
100 &  & \textbf{0.0273} & \underline{0.0227} & \textbf{0.3245} & \underline{0.4520} \\
\bottomrule

\end{tabular}
\end{center}
\vspace{-20pt}
\end{table}

In this section, we conducted benchmark experiments to verify the effectiveness and facilitate the use of Solid State Bus-Comp. We also conducted ablation studies on different data subsets to illustrate the effectiveness of improved data scale and diversity.

\subsection{Experiment Setup}

\quad \hskip0.6em\relax \textbf{Data Split and Processing}: For the train and evaluation data split, we randomly selected 112 songs as the train set and used the remaining 63 songs as the test set. We used our manually selected 144 parameter combinations for training and the seen test distribution. The remaining 76 parameter combinations are used as the unseen test distribution to assess the generalization ability.

\textbf{Training Schedules}: All the models are trained using the AdamW~\cite{adamw} optimizer with $\beta _ {1} = 0.9$, $\beta _ {2} = 0.999$, and a initial learning rate of 0.005. The ReduceLROnPlateau Scheduler is used with a factor of 0.5 and a patience of 10000 steps. All the experiments are conducted on a single NVIDIA H200 GPU with a batch size 16 and num workers of 16 for 500K steps. We use the Truncated Backpropagation Through Time (TBPTT)~\cite{tbptt} with a 0.01s segment length (4410 samples) to reduce memory costs while maintaining long-term dependencies.

\textbf{Baselines and Configurations}: We use the NablAFx~\cite{nablafx} toolbox for conducting benchmarking experiments on baseline systems. Specifically,  we use LSTM~\cite{LSTM-2019}, TCN~\cite{TCN-2021}, GCN~\cite{GCN-2023}, and S4~\cite{S4-2024} for black-box models. The LSTM model is conditioned on direct concatenation (Concat) or time-varying concatenation (TVConcat)~\cite{tonetwist}. The TCN, GCN, and S4 models are conditioned on FiLM~\cite{film}, temporal FiLM (TFiLM)~\cite{tfilm}, tiny temporal FiLM (TTFiLM)~\cite{tonetwist}, and time-varying temporal FiLM (TVFiLM)~\cite{tonetwist}. We use GreyBoxDRC~\cite{DRC-2022} and two compressor simulation chains proposed in ToneTwist~\cite{tonetwist} for grey-box models with the original configurations. For commercial plugins, we utilize the available models from Solid State Logic\footnote{\url{https://store.solidstatelogic.com/plug-ins/ssl-native-bus-compressor-2}}, Softube\footnote{\url{https://www.softube.com/bus-processor}}, Overloud\footnote{\url{https://www.overloud.com/products/comp-g}}, and PSPaudioware\footnote{\url{https://www.pspaudioware.com/products/psp-busspressor}}. To facilitate reproducible research, all of the modified code and the pre-trained models can be accessed via~\footnote{\url{https://drive.google.com/drive/folders/1zf5hnF7XGRW-poo_cqjQthKBeAZx33gd}}.

\textbf{Evaluation Metrics}: We use the Amphion~\cite{amphion} toolkit for objective evaluation. We use the L1 and Multi-Resolution STFT losses to evaluate the time and frequency-domain errors following ToneTwist~\cite{tonetwist}. We additionally report the number of trainable parameters to show the model size.

\subsection{Black-Box Methods}

Table~\ref{tab:results-black-box} illustrates the benchmarking results on black-box methods. Several key observations can be made: 1) Regarding the effectiveness of parameter scaling, LSTM and TCN models consistently benefit from increased model size. In contrast, GCN and S4 models only improve when conditioned on TTFiLM or TVFiLM layers. We speculate that the baseline FiLM layers used in these models are not expressive enough, leading to degraded performance as model capacity increases. On the other hand, TFiLM is powerful but introduces too many parameters, which may cause training instability in larger models. 2) Regarding parameter efficiency across different models, the LSTM model with TVConcat at 8.0K parameters achieves competitive results compared to larger models, and the S4 models with TVFiLM and TTFiLM reach near SOTA performance under 12K parameters. In contrast, models using TFiLM layers often require significantly more parameters to achieve comparable performance, making them unsuited for resource-constrained environments. 3) Regarding different conditioning layers, LSTM models with TVConcat perform significantly better than with simple concatenation. For TCN, GCN, and S4 models, TVFiLM surprisingly achieves the best performance, highlighting the effectiveness of time-varying modulation in modeling analog compressors. TFiLM generally ranks second, followed closely by TTFiLM, which offers a favorable trade-off between performance and parameter efficiency. 4) Regarding different model types, GCN consistently outperforms other architectures, demonstrating the strength of WaveNet-style dilated convolutions. S4 and TCN models with TTFiLM or TVFiLM also perform well. Notably, LSTM models with TVConcat outperform many other baselines, emphasizing the importance of temporal conditioning. 5) Regarding the generalization ability to unseen test scenarios, LSTM models with TVConcat and TCN, GCN, and S4 models with TFiLM, TTFiLM, and TVFiLM maintain strong performance on both seen and unseen parameter settings. In contrast, models using simpler conditioning layers exhibit noticeable performance drops under unseen testing senarios. 

\subsection{Grey-Box Methods}

The benchmarking results on grey-box models are presented in Table~\ref{tab:results-grey-box}. It can be observed that: 1) Regarding different gain computer models, static gain with a soft knee generally performs better with different level detectors. This aligns with the analog design of the SSL G-Bus compressor, which employs a soft knee where the knee width is automatically computed based on the threshold and ratio~\footnote{\label{mannal}\href{https://www.solidstatelogic.com/assets/uploads/downloads/SSL_500_Series_G_Comp_Module_User_Guide.pdf}{https://www.solidstatelogic.com/assets}}. 2) For different level detector implementations, the switching one-pole filter achieves the best overall performance, followed by the standard one-pole filter. In contrast, the RNN-modulated one-pole filter performs worse. We speculate that this is due to the relatively simple design of the VCA compressor's level detection circuit, which is different from the LA-2A that has strong non-linear distortion due to its optical components. Under this scenario, overly complex models like RNN-based detectors may overfit and lead to quality degradation. 3) egarding different test sets, a noticeable performance gap is observed between seen and unseen parameter settings. This can also be attributed to the changing compressor curve in the analog module~\footnoteref{mannal}, making it hard for grey-box models without explicit feedback mechanisms to capture that information. 4) In analog effect chain simulation, it is notable that the best performance is achieved using two parametric gain modules for compression and two parametric EQs for non-linear coloration. This illustrates the powerful learning ability of neural networks in loose conditions. Experiments also show adding a simple phase inversion module would damage the model performance since there are no phasers in the actual analog module, confirming its effectiveness and explainability.

\subsection{White-Box Plugins}

To evaluate the development of NN-based models and further illustrate the effectiveness of our proposed dataset, benchmarking results on white-box plugins are also reported, as shown in Table~\ref{tab:results-plugins}. Compared to these industry-standard plugins, a significant performance gap remains, particularly under extreme compression scenarios. This highlights that even the SOTA academic NN-based models still lag behind their commercial counterparts, which also illustrates the importance of our work since both model structure and datasets need to be improved for better performance.

\subsection{Ablation Study}

We also conducted ablation studies to illustrate the effectiveness of improving data quantity and diversity. We selected the GCN model conditioned with the TVFiLM layer as the baseline model and compared its performance when trained on different subsets. In particular, to control the data quantity, we fixed the number of total songs to 100 and control the length used to clip each song, resulting in 5 subsets from 3 minutes to 500 hours; to investigate the data diversity, we fixed the total data quantity to 50 hours and control the number of total songs with the adjusted clip lengths, resulting in 5 subsets from 5 songs to 100 songs. The detailed results are illustrated in Table~\ref{tab:results-ablation}. It can be observed that 1) increasing the data quantity steadily improves the model performance from 3 minutes to 500 hours, with the 50 hours as the division line for significant improvement, which is also confirmed by previous works~\cite{lauri-data}. 2) Increasing the data diversity is effective when there are only a few songs, and the improvement will be saturated until there are 50 different songs, especially in the unseen parameter settings.

\section{Conclusion}

In conclusion, this paper presents Solid State Bus-Comp, the first extensive and diverse dataset for DRC VA modeling. Our dataset comprises 2528 hours of processed unmastered songs in 220 parameter combinations with diverse genres, instruments, tempos, and keys. We provide benchmarking results on various open-sourced black-box and grey-box models, as well as available white-box plugins to facilitate the use of our dataset. We also provide ablation experiment results on different data subsets to illustrate the effectiveness of the improved data scale and quantity.

\section{Acknowledgment}

We acknowledge the computational resources provided by the Aalto Science-IT project. We acknowledge the EuroHPC Joint Undertaking for awarding this project access to the EuroHPC supercomputer LUMI, hosted by CSC (Finland) and the LUMI consortium through a EuroHPC Regular Access call. This work is also supported by the 2023 Shenzhen stability Science Program, the Program for Guangdong Introducing Innovative and Entrepreneurial Teams (Grant No. 2023ZT10X044), and the Shenzhen Science and Technology Program (ZDSYS20230626091302006)


\nocite{*}
\bibliographystyle{IEEEbib}
\bibliography{DAFx25_tmpl} 

\begin{thebibliography}{10}

\bibitem{drc-review-2012}
Dimitrios Giannoulis, Michael Massberg, and Joshua~D Reiss,
\newblock ``{Digital dynamic range compressor design—A tutorial and analysis},''
\newblock {\em {Journal of the Audio Engineering Society}}, vol. 60, no. 6, pp. 399--408, 2012.

\bibitem{review-1984}
Guy~W McNally,
\newblock ``{Dynamic range control of digital audio signals},''
\newblock {\em {Journal of the Audio Engineering Society}}, vol. 32, no. 5, pp. 316--327, 1984.

\bibitem{block-drc}
Germ{\'a}n Ramos,
\newblock ``{Block processing strategies for computationally efficient dynamic range controllers},''
\newblock in {\em {Proc. Int. Conf. Digital Audio Effects}}, 2011, pp. 253--256.

\bibitem{FFT-drc}
Leo McCormack and Vesa V{\"a}lim{\"a}ki,
\newblock ``{FFT-based dynamic range compression},''
\newblock in {\em {Sound Music Comput. Conf.}}, 2017, pp. 42--49.

\bibitem{automatic-2013}
Dimitrios Giannoulis, Michael Massberg, and Joshua~D Reiss,
\newblock ``{Parameter automation in a dynamic range compressor},''
\newblock {\em {Journal of the Audio Engineering Society}}, vol. 61, no. 10, pp. 716--726, 2013.

\bibitem{multitrack-2013}
Jacob~A Maddams, Saoirse Finn, and Joshua~D Reiss,
\newblock ``{An autonomous method for multi-track dynamic range compression},''
\newblock in {\em {Proc. Int. Conf. Digital Audio Effects}}, 2012, pp. 1--8.

\bibitem{device-2011}
Oliver Kr{\"o}ning, Kristjan Dempwolf, and Udo Z{\"o}lzer,
\newblock ``{Analysis and simulation of an analog guitar compressor},''
\newblock {\em {Proc. Int. Conf. Digital Audio Effects}}, pp. 205--208, 2011.

\bibitem{optocoupler-2016}
Felix Eichas and Udo Z{\"o}lzer,
\newblock ``{Modeling of an optocoupler-based audio dynamic range control circuit},''
\newblock in {\em {Novel Optical Systems Design and Optimization XIX}}, 2016, vol. 9948, pp. 47--62.

\bibitem{data-2023}
Alessandro~Ilic Mezza, Riccardo Giampiccolo, and Alberto Bernardini,
\newblock ``{Data-Driven Parameter Estimation of Lumped-Element Models via Automatic Differentiation},''
\newblock {\em {IEEE} Access}, vol. 11, pp. 143601--143615, 2023.

\bibitem{la2a}
Scott~H Hawley, Benjamin Colburn, and Stylianos~I Mimilakis,
\newblock ``{SignalTrain: Profiling audio compressors with deep neural networks},''
\newblock {\em arXiv:1905.11928}, 2019.

\bibitem{LSTM-2019}
Alec Wright, Eero-Pekka Damsk{\"a}gg, and Vesa V{\"a}lim{\"a}ki,
\newblock ``{Real-time black-box modelling with recurrent neural networks},''
\newblock in {\em {Proc. Int. Conf. Digital Audio Effects}}, 2019.

\bibitem{hrnn}
Yen-Tung Yeh, Wen-Yi Hsiao, and Yi-Hsuan Yang,
\newblock ``{Hyper recurrent neural network: Condition mechanisms for black-box audio effect modeling},''
\newblock {\em arXiv:2408.04829}, 2024.

\bibitem{wavenet-audio}
Alec Wright, Eero-Pekka Damsk{\"a}gg, Lauri Juvela, and Vesa V{\"a}lim{\"a}ki,
\newblock ``{Real-time guitar amplifier emulation with deep learning},''
\newblock {\em {Applied Sciences}}, vol. 10, no. 3, pp. 766, 2020.

\bibitem{wavenet}
Aaron Van Den~Oord, Sander Dieleman, Heiga Zen, Karen Simonyan, Oriol Vinyals, Alex Graves, Nal Kalchbrenner, Andrew Senior, Koray Kavukcuoglu, et~al.,
\newblock ``{Wavenet: A generative model for raw audio},''
\newblock {\em arXiv:1609.03499}, 2016.

\bibitem{TCN-2021}
Christian~J Steinmetz and Joshua~D Reiss,
\newblock ``{Efficient neural networks for real-time modeling of analog dynamic range compression},''
\newblock {\em arXiv:2102.06200}, 2021.

\bibitem{GCN-2023}
Marco Comunit{\`a}, Christian~J Steinmetz, Huy Phan, and Joshua~D Reiss,
\newblock ``{Modelling black-box audio effects with time-varying feature modulation},''
\newblock in {\em {Proc. IEEE Int. Conf. Acoustics, Speech and Signal Processing}}, 2023, pp. 1--5.

\bibitem{film}
Ethan Perez, Florian Strub, Harm de~Vries, Vincent Dumoulin, and Aaron~C. Courville,
\newblock ``{FiLM: Visual Reasoning with a General Conditioning Layer},''
\newblock in {\em {Proc. AAAI Conf. Artif. Intell.}}, 2018, vol.~32.

\bibitem{ssm-2022}
Albert Gu, Karan Goel, and Christopher R{\'{e}},
\newblock ``{Efficiently Modeling Long Sequences with Structured State Spaces},''
\newblock in {\em {Proc. Int. Conf. Learn. Representations}}, 2022.

\bibitem{S4-2024}
Hanzhi Yin, Gang Cheng, Christian~J Steinmetz, Ruibin Yuan, Richard~M Stern, and Roger~B Dannenberg,
\newblock ``{Modeling analog dynamic range compressors using deep learning and state-space models},''
\newblock {\em arXiv:2403.16331}, 2024.

\bibitem{S6-2024}
Riccardo Simionato and Stefano Fasciani,
\newblock ``{Comparative study of recurrent neural networks for virtual analog audio effects modeling},''
\newblock {\em arXiv:2405.04124}, 2024.

\bibitem{mamba}
Albert Gu and Tri Dao,
\newblock ``{Mamba: Linear-time sequence modeling with selective state spaces},''
\newblock {\em arXiv:2312.00752}, 2023.

\bibitem{ddsp}
Jesse~H. Engel, Lamtharn Hantrakul, Chenjie Gu, and Adam Roberts,
\newblock ``{DDSP: Differentiable Digital Signal Processing},''
\newblock in {\em {Proc. Int. Conf. Learn. Representations}}, 2020.

\bibitem{biquad-2020}
Boris Kuznetsov, Julian~D Parker, and Fabi{\'a}n Esqueda,
\newblock ``{Differentiable IIR filters for machine learning applications},''
\newblock in {\em {Proc. Int. Conf. Digital Audio Effects}}, 2020, pp. 297--303.

\bibitem{klann-2024}
Ville Huhtala, Lauri Juvela, and Sebastian~J. Schlecht,
\newblock ``{KLANN: Linearising Long-Term Dynamics in Nonlinear Audio Effects Using Koopman Networks},''
\newblock {\em {IEEE Signal Process. Lett.}}, vol. 31, pp. 1169--1173, 2024.

\bibitem{koopman-network}
Bethany Lusch, J~Nathan Kutz, and Steven~L Brunton,
\newblock ``{Deep learning for universal linear embeddings of nonlinear dynamics},''
\newblock {\em {Nature Communications}}, vol. 9, no. 1, pp. 4950, 2018.

\bibitem{DRC-2022}
Alec Wright and Vesa Valimaki,
\newblock ``{Grey-box modelling of dynamic range compression},''
\newblock in {\em {Proc. Int. Conf. Digital Audio Effects}}, 2022, pp. 304--311.

\bibitem{nablafx}
Marco Comunit{\`a}, Christian~J Steinmetz, and Joshua~D Reiss,
\newblock ``{NablAFx: A Framework for Differentiable Black-box and Gray-box Modeling of Audio Effects},''
\newblock {\em arXiv:2502.11668}, 2025.

\bibitem{UA6176}
Marco~A Mart{\'\i}nez~Ram{\'\i}rez, Emmanouil Benetos, and Joshua~D Reiss,
\newblock ``{Deep learning for black-box modeling of audio effects},''
\newblock {\em {Applied Sciences}}, vol. 10, no. 2, pp. 638, 2020.

\bibitem{cl1b}
Riccardo Simionato and Stefano Fasciani,
\newblock ``{Fully conditioned and low-latency black-box modeling of analog compression},''
\newblock in {\em {Proc. Int. Conf. Digital Audio Effects}}, 2023.

\bibitem{mert}
Yizhi Li, Ruibin Yuan, Ge~Zhang, Yinghao Ma, Xingran Chen, Hanzhi Yin, Chenghao Xiao, Chenghua Lin, Anton Ragni, Emmanouil Benetos, Norbert Gyenge, Roger~B. Dannenberg, Ruibo Liu, Wenhu Chen, Gus Xia, Yemin Shi, Wenhao Huang, Zili Wang, Yike Guo, and Jie Fu,
\newblock ``{MERT: Acoustic Music Understanding Model with Large-Scale Self-supervised Training},''
\newblock in {\em {Proc. Int. Conf. Learn. Representations}}, 2024.

\bibitem{stableaudio}
Zach Evans, Julian~D Parker, CJ~Carr, Zack Zukowski, Josiah Taylor, and Jordi Pons,
\newblock ``{Stable audio open},''
\newblock in {\em {Proc. IEEE Int. Conf. Acoustics, Speech and Signal Processing}}, 2025, pp. 1--5.

\bibitem{maskgct}
Yuancheng Wang, Haoyue Zhan, Liwei Liu, Ruihong Zeng, Haotian Guo, Jiachen Zheng, Qiang Zhang, Xueyao Zhang, Shunsi Zhang, and Zhizheng Wu,
\newblock ``{MaskGCT: Zero-shot text-to-speech with masked generative codec transformer},''
\newblock in {\em {Proc. Int. Conf. Learn. Representations}}, 2024.

\bibitem{singnet}
Yicheng Gu, Chaoren Wang, Junan Zhang, Xueyao Zhang, Zihao Fang, Haorui He, and Zhizheng Wu,
\newblock ``{SingNet: Towards a Large-Scale, Diverse, and In-the-Wild Singing Voice Dataset},''
\newblock {\em {OpenReview}}, 2024.

\bibitem{neurodyne}
Yicheng Gu, Chaoren Wang, Zhizheng Wu, and Lauri Juvela,
\newblock ``{Neurodyne: Neural Pitch Manipulation with Representation Learning and Cycle-Consistency GAN},'' 2025.

\bibitem{yue}
Ruibin Yuan, Hanfeng Lin, Shuyue Guo, Ge~Zhang, Jiahao Pan, Yongyi Zang, Haohe Liu, Yiming Liang, Wenye Ma, Xingjian Du, et~al.,
\newblock ``{YuE: Scaling Open Foundation Models for Long-Form Music Generation},''
\newblock {\em arXiv:2503.08638}, 2025.

\bibitem{emilia}
Haorui He, Zengqiang Shang, Chaoren Wang, Xuyuan Li, Yicheng Gu, Hua Hua, Liwei Liu, Chen Yang, Jiaqi Li, Peiyang Shi, Yuancheng Wang, Kai Chen, Pengyuan Zhang, and Zhizheng Wu,
\newblock ``{Emilia: An Extensive, Multilingual, and Diverse Speech Dataset for Large-Scale Speech Generation},''
\newblock in {\em {Proc. IEEE Spoken Lang. Technol. Workshop}}, 2024, pp. 885--890.

\bibitem{emilia-journal}
Haorui He, Zengqiang Shang, Chaoren Wang, Xuyuan Li, Yicheng Gu, Hua Hua, Liwei Liu, Chen Yang, Jiaqi Li, Peiyang Shi, et~al.,
\newblock ``{Emilia: A Large-Scale, Extensive, Multilingual, and Diverse Dataset for Speech Generation},''
\newblock {\em arXiv:2501.15907}, 2025.

\bibitem{vevo}
Xueyao Zhang, Xiaohui Zhang, Kainan Peng, Zhenyu Tang, Vimal Manohar, Yingru Liu, Jeff Hwang, Dangna Li, Yuhao Wang, Julian Chan, et~al.,
\newblock ``{Vevo: Controllable zero-shot voice imitation with self-supervised disentanglement},''
\newblock in {\em {Proc. Int. Conf. Learn. Representations}}, 2025.

\bibitem{tonetwist}
Marco Comunit{\`a}, Christian~J Steinmetz, and Joshua~D Reiss,
\newblock ``{Differentiable Black-box and Gray-box Modeling of Nonlinear Audio Effects},''
\newblock {\em arXiv:2502.14405}, 2025.

\bibitem{tempocnn}
Hendrik Schreiber and Meinard M{\"u}ller,
\newblock ``{Musical tempo and key estimation using convolutional neural networks with directional filters},''
\newblock {\em arXiv:1903.10839}, 2019.

\bibitem{qwen2audio}
Yunfei Chu, Jin Xu, Qian Yang, Haojie Wei, Xipin Wei, Zhifang Guo, Yichong Leng, Yuanjun Lv, Jinzheng He, Junyang Lin, et~al.,
\newblock ``{Qwen2-audio technical report},''
\newblock {\em arXiv:2407.10759}, 2024.

\bibitem{llama3}
Aaron Grattafiori, Abhimanyu Dubey, Abhinav Jauhri, Abhinav Pandey, Abhishek Kadian, Ahmad Al-Dahle, Aiesha Letman, Akhil Mathur, Alan Schelten, Alex Vaughan, et~al.,
\newblock ``{The llama 3 herd of models},''
\newblock {\em arXiv:2407.21783}, 2024.

\bibitem{w2vbert}
Yu{-}An Chung, Yu~Zhang, Wei Han, Chung{-}Cheng Chiu, James Qin, Ruoming Pang, and Yonghui Wu,
\newblock ``{w2v-BERT: Combining Contrastive Learning and Masked Language Modeling for Self-Supervised Speech Pre-Training},''
\newblock in {\em {Proc. IEEE Autom. Speech Recognit. Understanding Workshop}}, 2021, pp. 244--250.

\bibitem{adamw}
Ilya Loshchilov and Frank Hutter,
\newblock ``{Decoupled Weight Decay Regularization},''
\newblock in {\em {Proc. Int. Conf. Learn. Representations}}, 2019.

\bibitem{tbptt}
Christopher Aicher, Nicholas~J. Foti, and Emily~B. Fox,
\newblock ``{Adaptively Truncating Backpropagation Through Time to Control Gradient Bias},''
\newblock in {\em {Conf. Uncertain. Artif. Intell.}}, 2019, pp. 799--808.

\bibitem{tfilm}
Marco Comunit{\`{a}}, Christian~J. Steinmetz, Huy Phan, and Joshua~D. Reiss,
\newblock ``{Modelling Black-Box Audio Effects with Time-Varying Feature Modulation},''
\newblock in {\em {Proc. IEEE Int. Conf. Acoustics, Speech and Signal Processing}}, 2023, pp. 1--5.

\bibitem{amphion}
Xueyao Zhang, Liumeng Xue, Yicheng Gu, Yuancheng Wang, Jiaqi Li, Haorui He, Chaoren Wang, Ting Song, Xi~Chen, Zihao Fang, Haopeng Chen, Junan Zhang, Tze~Ying Tang, Lexiao Zou, Mingxuan Wang, Jun Han, Kai Chen, Haizhou Li, and Zhizheng Wu,
\newblock ``{Amphion: An Open-Source Audio, Music and Speech Generation Toolkit},''
\newblock in {\em {Proc. IEEE Spoken Lang. Technol. Workshop (SLT)}}, 2024, pp. 879--884.

\bibitem{lauri-data}
Lauri Juvela, Eero-Pekka Damsk{\"a}gg, Aleksi Peussa, Jaakko M{\"a}kinen, Thomas Sherson, Stylianos~I Mimilakis, Kimmo Rauhanen, and Athanasios Gotsopoulos,
\newblock ``{End-to-end amp modeling: from data to controllable guitar amplifier models},''
\newblock in {\em {Proc. IEEE Int. Conf. Acoustics, Speech and Signal Processing}}, 2023, pp. 1--5.

\end{thebibliography}

\end{document}